\documentclass[11pt]{llncs}
\usepackage{amsmath}
\usepackage{bbm}
\usepackage{amsfonts}
\usepackage{epic,eepic,latexsym,verbatim}

\newcommand{\ket}[1]{|#1\rangle}
\newcommand{\bra}[1]{\langle #1|}
\newcommand{\bracket}[2]{\langle #1|#2\rangle}
\newcommand{\ketbra}[1]{|#1\rangle\langle #1|}
\newcommand{\cket}[1]{|\widetilde{#1}\rangle}

\newcommand{\bk}[0]{{\bf k}}
\newcommand{\bx}{{\bf x}}
\usepackage{amssymb}
\begin{document}

\title{Optimal State Merging Without Decoupling}
\author{Jean-Christian Boileau\inst{1} 
\and 
Joseph M.~Renes\inst{2}
}

\institute{Center for Quantum Information and Quantum Control, University of Toronto
\and
Institut f\"ur Angewandte Physik, Technical University of Darmstadt
}

\maketitle       

\begin{abstract}
We construct an optimal state merging protocol by adapting a recently-discovered optimal 
entanglement distillation protcol [Renes and Boileau, Phys.\ Rev.\ A .\ 73, 032335 (2008)]. The proof of optimality relies only on directly
establishing sufficient ``amplitude'' and ``phase'' correlations between Alice and Bob and not on usual techniques of decoupling Alice
from the environment. This strengthens the intuition from quantum error-correction that these two correlations are all that
really matter in two-party quantum information processing. 
\end{abstract}

\section{Introduction}
Quantum state merging is an important primitive protocol in the hierarchy of quantum communication protocols, also known as 
the quantum information family tree. Given two parties Alice and Bob and a mixed bipartite state $\psi^{AB}$, the goal of state merging is 
simply for Alice to send her half of the state to Bob. One option, of course, is to compress $\psi^A$ into as few qubits as possible and 
send it over a quantum channel. However, this ignores the information Bob has about the state in the form of $\psi^B$. 
Although it might seem that a quantum channel is essential for state merging to work, Bob's side information can be such that only
\emph{classical} communication from Alice is required. 

Reasoning about the protocol is made somewhat easier  by considering the purification $\ket{\psi}^{ABR}$ of $\psi^{AB}$ 
to a reference system $R$, 
that is $\psi^{AB}={\rm Tr}_R\left[\psi^{ABR}\right]$. 
The goal of state merging is then to arrange for Bob to hold the purification of $R$. In some cases quantum communication will clearly be required, 
for instance when $\ket{\psi}^{ABR}=\ket{\Phi}^{AR}\ket{\xi}^{B}$, 
where $\ket{\xi}$ is arbitrary while $\ket{\Phi}^{AR}=\frac{1}{\sqrt{d}}\sum_k \ket{kk}^{AR}$ is the canonical maximally entangled state 
for a fixed basis $\{\ket{k}\}$ and $d$ is the minimum dimension of $A$ and $R$. 
Bob's state is clearly irrelevant, and Alice must simply send her whole system, as it is incompressible.  
On the other hand, when Alice and Bob share $\ket{\Phi}^{AB}$, no
communication is required at all! 
This is simply due to the fact that now the state of $R$ is by itself pure, so neither Alice nor Bob hold its purification. 

Horodecki, Oppenheim, and Winter~\cite{horodecki_partial_2005,horodecki_quantum_2007} consider the asymptotic setting of 
many copies of $\psi^{ABR}$ and show that classical communication suffices when the quantum conditional entropy $S(A|B)=S(AB)-S(B)$ is negative,
where $S(A)=-{\rm Tr}\left[\rho^A\log_2\rho^A\right]$ is the von Neumann entropy. 
In fact, when $S(A|B)<0$ their state merging 
protocol {produces} entangled pairs at the rate $-S(A|B)$ and uses classical communication at the rate $I(A{:}R)$, 
where $I(A{:}R)=S(A)+S(R)-S(AR)$ is the quantum 
mutual information. These rates are also shown to be optimal.  
When $S(A|B)>0$ on the other hand, any state merging protocol requires quantum communication at the rate $S(A|B)$, 
or equivalently consumes entangled pairs at this rate.   
This fact gives an operational meaning to the conditional entropy in terms of entanglement consumption or production, which due to its possible negativity is quite unlike its classical counterpart.  

In this paper we construct a state merging protocol operating at the optimal rates by focusing 
on the classical information that Bob has about 
complementary observables ``amplitude'' and ``phase'' on Alice's system and showing how classical communication is sufficient to transfer 
the necessary quantum correlations. 
This approach is substantially different from the original proof, which is based on the technique of decoupling Alice's system
from the reference system $R$~\cite{schumacher_approximate_2002},
and follows our recent work on entanglement distillation (ED) quite closely~\cite{renes_physical_2008}. Indeed, 
state merging is actually achieved in that protocol as well, but at the cost of too much classical communication. We rectify this problem here, showing
that if Alice first compresses her system and then runs the ED protocol, a small modification suffices to make this an optimal state merging protocol. 

The remainder of the paper is outlined as follows. We first review the known results for the state merging protocol in the next section, and
then recapitulate the important parts of the proof of the ED protocol appearing in~\cite{renes_physical_2008} in the following section. Section~\ref{sec:SMmod} contains the new contribution of this paper, showing how to modify the ED protocol to use only the 
minimum necessary classical communication. 
Finally, we conclude with a summary of the results and comment on the connections to the quantum noisy channel coding theorem.  

\section{\label{sec:SMrevisit} State Merging Defined}
As with most protocols in quantum information theory, we are concerned here with the rate at which Alice and Bob can transform 
an asymptotically-large number of copies of the state $\ket{\psi}^{ABR}$ into a good approximation of $n$ copies in which Bob holds system $A$. 
To keep the accounting simple, we assume that any necessary quantum communication is performed by teleportation through pre-shared entangled pairs, so that
the protocol uses only classical communication in any case, and either produces or consumes entanglement depending on the circumstances. 
We then define an $(n,\epsilon)$ state merging protocol for $\psi^{ABR}$ to be a series
of local operations involving only classical communication (LOCC operations) such that application to $\ket{\Psi}^{ABR}=(\ket{\psi}^{ABR})^{\otimes n}$
produces an output ${\Upsilon}^{DBR}$ in which Bob holds the system $D$ such that $||{\Upsilon}^{DBR}-\Psi^{DBR}||_1\leq \epsilon$. 
If there exists an $(n,\epsilon_n)$ protocol using $K_n$ bits of classical communication and consuming $E_n$ ebits of entanglement for every $n$ such that $\lim_{n\rightarrow\infty}\epsilon_n=0$, then the rates of communication and entanglement consumption of the protocol are given by 
\begin{align}
R_K=\lim_{n\rightarrow\infty}\frac{K_n}{n}\qquad \text{and}\qquad R_E=\lim_{n\rightarrow\infty}\frac{E_n}{n}.
\end{align}
Horodecki, Oppenheim, and Winter showed in~\cite{horodecki_partial_2005,horodecki_quantum_2007} that 
\begin{align}
\label{eq:rates}
\inf R_K= I(A{:}R)\qquad \text{and}\qquad \inf R_E= S(A|B),
\end{align}
where a negative $R_E$ indicates the amount of entanglement produced. The proof of these statements has two parts, the direct part showing the
rates are achievable, and the converse part showing they cannot be surpassed. Here we will give a new proof of the direct part, borrowing our  
techniques from~\cite{renes_physical_2008} which were used to give a new proof of the hashing inequality~\cite{devetak_distillation_2005} on the achievable rate of entanglement 
distillation. In the next section we sketch the important parts of that proof.

\section{\label{sec:EDproof} Entanglement Distillation Revisited}
A maximally entangled pair in one for which Bob can predict the measurement of either of the two observables, ``amplitude'' $Z^A=\sum_k (-1)^{k}\ketbra{k}^A$ and its Fourier conjugate ``phase''
$X=\sum_k \ket{k\oplus 1}\bra{k}$. Here we are assuming that Alice's system has dimension 2, but what follows can be easily extended to higher dimensions. 
Since this is the desired output of the distillation procedure, the 
idea behind the protocol given in Theorem 6 of~\cite{renes_physical_2008} is to determine what information Bob already has about these observables from his system $B$
and then arrange for Alice to send him the rest. This is classical information, since it refers to the measurement outcomes, and therefore only
classical communication will be required. However, since Alice needs to send information pertaining to both $X$ and $Z$, one must ensure 
that both parts of her message simultaneously exist. This is achieved by measuring the $X$- and $Z$-type stabilizers of a
Calderbank-Shor-Steane (CSS) code~\cite{calderbank_good_1996,steane_multiple-particle_1996,nielsen_quantum_2000}
to generate the message. The amount of
information is governed by the ``static'' version of the Holevo-Schumacher-Westmoreland (HSW) theorem~\cite{holevo_capacity_1998,schumacher_sending_1997}, 
which we review in the appendix. 

Greatly simplified, 
the protocol starts by Alice picking a random CSS code of a given size for her Hilbert space. She then 
measures the stabilizers to obtain the syndromes ${\bf \alpha}$ (for $X$) and ${\bf \beta}$ (for $Z$) and communicates them to Bob. 
The syndromes are such that he can find measurements $\Lambda_{\bf \alpha,x}^B$ and $\Gamma_{\bf \beta,z}^B$ on $B$ which enable him to predict (with 
high probability) the 
outcome of measuring either $X^A$ or $Z^A$, respectively. The existence of such measurements is guaranteed by the (static) HSW theorem, using
Bob's marginal states generated by Alice's measurement as the ensemble and the code syndrome as the side information. 
It implies that the CSS code must have roughly $m_Z=nS(Z^A|B)$ $Z$-type syndromes and 
$m_X=nS(X^A|CB)$ $X$-type, where $C$ is an additional quantum register containing a copy of Alice's system in the $Z$ basis, and
$S(Z^A|B)=S(\bar{\psi}_Z^{AB})-S(\psi^B)$ for $\bar{\psi}_Z^{AB}$ the shared state after Alice measures the observable $Z$. Once this
process is complete, Bob can (in principle) predict either $X^A$ or $Z^A$ on each pair, and therefore 
can perform a quantum operation on his systems to create entangled pairs (to good approximation). Since Alice is left 
with only the code subspace given by $\alpha$ and $\beta$, whose size is $n-m_X-m_X$, 
this is the number of entangled pairs they can create.  

To see how this works in more detail, begin with the individual shared state $\ket{\psi}^{ABR}$ and write it as
$\ket{\psi}^{ABR}=\sum\sqrt{p_k}\ket{k}^A\ket{\varphi_k}^{BR}$,
where $\ket{k}$ is the eigenbasis of $\psi^A$ and also defines the operator $Z$, the $\ket{\varphi_k}$ are a set of arbitrary orthonormal states, and $p_k$ is a probability distribution. 
The $n$-fold version 
 $\ket{\Psi_0}^{ABR}=(\ket{\psi}^{ABR})^{\otimes n}$ we write like so, using bold-faced symbols $\bk$ to denote strings $(k_1,k_2,\dots,k_n)$:
\begin{align}
\ket{\Psi_0}^{ABR}=\sqrt{p_\bk}\sum_{\bk}\ket{\bk}^A\ket{\varphi_{\bk}}^{BR}.
\end{align}  
We'll also need to consider the associated state in which Bob
has a copy of Alice's system in the $Z$ basis:
\[ \ket{\psi_c}^{ACBR}=\sum\sqrt{p_k}\ket{kk}^{AC}\ket{\varphi_k}^{BR}=\tfrac{1}{\sqrt{2}} \cket{x}^A\ket{\vartheta_x}^{CBR}.\]
Here $\cket{x}$ is an eigenstate of $X$
and the $\ket{\vartheta_x}$ are again a arbitrary set of orthonormal states. Observe that $\ket{\vartheta_0}^{CBR}=\ket{\psi}^{CBR}$. 

Denote the projections onto the stabilizers of the chosen CSS code by $\widetilde{\Pi}_\alpha^A$ and $\Pi_\beta^A$, which commute by the CSS 
nature of the code. The result of Alice measuring the stabilizers and sending them to Bob is
\begin{align}
\ket{\Psi_1}^{ABRP}=\sum_{\alpha,\beta}\widetilde{\Pi}_\alpha^A\Pi_\beta^A\ket{\Psi_0}^{ABR}\ket{\alpha,\beta}^P.
\end{align}
The system label $P$, for ``public'', is shorthand for having arbitrarily many copies $P_1, P_2,\dots$ of the values $\alpha,\beta$, and mimics the
information being classically-transmitted. Given $\beta$, Bob can coherently perform the measurement $\Gamma_{\bf \beta, k}^B$  to
extract the value of $k$ in $A$ to an auxiliary system $C$ with high probability. 
One can show that this implies the state is very nearly identical to  
\begin{align}
\ket{\Psi_2}&=\sum_{\alpha,\beta,\bk}\widetilde{\Pi}_\alpha^A\Pi_\beta^A\ket{\bk\bk}^{AC}\ket{\varphi_\bk}^{BR}\ket{\alpha,\beta}^P=\sum_{\alpha,\beta}\widetilde{\Pi}_\alpha^A\Pi_\beta^A\ket{\Psi_c}\ket{\alpha,\beta}^{P}.
\end{align}

Next, Bob can coherently measure $\Lambda_{\bf \alpha,x}^B$ to extract $\bx$ in the conjugate basis of $A$ to a further auxiliary system $D$, again with 
high probability. The resulting state is nearly identical to 
\begin{align}
\label{eq:psi3}
\ket{\Psi_3}&=\tfrac{1}{\sqrt{2^n}}\sum_{\alpha,\beta,\bx}\widetilde{\Pi}_\alpha^A\Pi_\beta^A\ket{\widetilde{\bx}}^A\ket{\widetilde{\bx}}^D\ket{\vartheta_\bx}^{CBR}\ket{\alpha,\beta}^P.
\end{align}
Owing to the properties of $X$ and $Z$ and the two forms of $\ket{\psi_c}$, we have the relation
$\ket{\vartheta_\bx}^{CBR}=\sum_\bk \sqrt{p_\bk}\,\omega^{\bk\cdot \bx}\ket{\bk}^C\ket{\varphi_\bk}^{BR}=(Z^\bx)^C\ket{\Psi_0}^{CBR}
$.
Inserting this into equation~\ref{eq:psi3} gives
\begin{align}
\ket{\Psi_3}&=\tfrac{1}{\sqrt{2^n}}\sum_{\alpha,\beta,\bx}\widetilde{\Pi}_\alpha^A\Pi_\beta^A\ket{\widetilde{\bx}}^A\ket{\widetilde{\bx}}^D(Z^\bx)^C\ket{\Psi_0}^{CBR}\ket{\alpha,\beta}^P.
\end{align}

Finally, a controlled-$Z$ operation from $D$ to $C$ inverts the $Z^\bx$ operator, leaving the desired output
\begin{align}
\ket{\Psi_4}&=\sum_{\alpha,\beta,\bx}\widetilde{\Pi}_\alpha^A\Pi_\beta^A\ket{\Phi_n}^{AD}\ket{\alpha,\beta}^P\otimes \ket{\Psi_0}^{CBR},
\end{align} 
where $\ket{\Phi_n}=\ket{\Phi}^{\otimes n}$. 
Observe that the purification of $R$ is now solely in Bob's possession, so state merging has been accomplished. Furthermore, since $n[S(Z^A|B)+S(X^A|CB)]$ CSS stabilizers leave $n[1-S(Z^A|B)-S(X^A|CB)]$ encoded logical operators, Alice and Bob share this many entangled pairs in systems $A$ and $D$. In~\cite{renes_physical_2008} it is shown that this equals $-nS(A|B)$, so provided this quantity is positive ($S(A|B)<0$), the protocol achieves the rate $R_E$. 

Of course, $\ket{\Psi_4}$ is not precisely the output of the protocol, since the two coherent measurement operations by Bob were not perfect. The 
details of the approximation are given in~\cite{renes_physical_2008}, the result being that if Alice chooses a random code having  
$n[S(Z^A|B)+\delta]$ $Z$-type stabilizers and $n[S(X^A|CB)+\delta]$ $X$-type stabilizers for some $\delta>0$, then the 
output will be within $\exp(-O(n\delta^2))$ 
of $\ket{\Psi_4}$, as measured by the trace-distance.

If $S(A|B)>0$, we can use the same trick as~\cite{horodecki_partial_2005,horodecki_quantum_2007}. Adding $n[S(A|B)+2\delta]$ entangled
pairs, each of which has $S(A|B)=-1$, the conditional entropy of the overall state $\ket{\Psi}^{ABR}\ket{\Phi_{n[S(A|B)+\epsilon]}}^{A'B'}$ is $-2n\delta$. Using this as the individual input into the above protocol accomplishes the state merging and outputs no entanglement. In this way $R_E$ can 
be achieved when $S(A|B)>0$.  

The above protocol requires too much classical communication, however, $n[1-S(A|B)]$ bits. This is
generally greater than $I(A{:}E)$, and is only equal for $S(A)=1$. The fact that the protocol is optimal when $\psi^A$ is maximally mixed
suggests that for a general input Alice should first compress her system and then run the protocol. However, the compression procedure will disturb the conjugate 
observable $X$ and its eigenbasis, so there is no longer any guarantee that Bob's $\Lambda_{\alpha,\bx}$ measurement will work as intended. 
The next section shows how to fix this problem.    

\section{\label{sec:SMmod} Classical Communication Reduced}

Fortunately, the ensemble of states $\vartheta_\bx^{CB}$ which Bob would like to distinguish is invariant under the action of the group
$(Z^\bx)^C$, which will enable us to adapt the original $\Lambda_{\alpha,\bx}$ measurement for use after Alice compresses her state. 
This will reduce the number of $X$ syndromes she needs to communicate to Bob to the optimal level. 

The modified  protocol begins as before with the state $\ket{\Psi_0}$. Alice then makes a measurement projecting her systen onto 
the typical subspace $\mathcal{T}_\delta^n$, which is the subspace spanned by eigenvectors $\ket{\bk}$ whose $\bk$ are in the
typical set $T_\delta^n=\{\bk:|-\frac{1}{n}\log p_\bk-S(\psi^A)|\leq \delta\}$ for a fixed $\delta>0$~\cite{schumacher_quantum_1995,cover_elements_2006}. 
The probability $\mathcal{N}_\delta^n={\rm Pr}[\bk\in T_\delta^n]$ that $\bk$ is typical is greater than $1-2^{-cn\delta^2}:=1-\epsilon$, for some 
constant $c$~\cite{devetak_distillation_2005}
and therefore the projection succeeds
with probability exponentially close to unity; otherwise the protocol aborts. When it succeeds, it \emph{prunes} the state $\ket{\Psi_0}$, 
leaving
\begin{align}
\ket{\Psi'_0}^{ABR}=\frac{1}{\sqrt{\mathcal{N}_\delta^n}}\sum_{\bk\in T_\delta^n}\sqrt{p_\bk}\ket{\bk}^A\ket{\varphi_\bk}^{BR}=\sum_{\bk\in T_\delta^n}\sqrt{p'_{\bk}}\ket{\bk}^A\ket{\varphi_{\bk}}^{BR},
\end{align}
where we have implicitly defined new probability weights $p'_\bk=p_\bk/\mathcal{N}_\delta^n$. 
Importantly, $D_\delta^n:={\rm dim}(\mathcal{T}_\delta^n)\leq 2^{n[S(\psi^A)+\delta]}$, and a simple calculation shows that $\bracket{\Psi_0}{\Psi_0'}=\sqrt{\mathcal{N}_\delta^n}$. This implies that two states are close in trace distance, $||\Psi_0-\Psi'_0||_1\leq \sqrt{\epsilon}$, using
the relationship between fidelity and trace distance $||\rho-\sigma||_1\leq \sqrt{1-F(\rho,\sigma)^2}$~\cite{fuchs_cryptographic_1999}.

The protocol proceeds just as before, measuring $X'$- and $Z'$-type stabilizers of a random CSS code on the 
pruned state and communicating the results to Bob. Here $Z'$ is the analog of $Z$ for the typical subspace, and $X'$ is its Fourier conjugate.  
Now, however, we have no direct way of setting the number of stabilizers, since the state is no longer i.i.d.\ and therefore the HSW theorem
no longer applies. 
This is not really a problem for the $Z'$-type stabilizers, since the typical projection is done in the $\ket{\bk}$ basis, the basis
which generates the $\varphi_\bk^B$. 
By design, the measurement constructed in the HSW theorem
does not attempt to identify $\varphi_\bk^B$ for nontypical $\bk$, so Bob can just reuse it in this case. The probability of error will 
only decrease by \emph{explicitly} rejecting nontypical $\bk$. Hence $m_z\approx nS(Z^A|B)$ as before. 

However, the original measurement will not work for the conjugate basis $\ket{\bx'}$, 
the Fourier transform of the typical subspace basis, since the states 
$\vartheta'^{CB}_{\bx'}$ have no \emph{a priori} relation to the original $\vartheta_{\bx}^{CB}$. However, the former states stem from the related state
\begin{align}
\ket{\Psi_c'}=\sum_{\bk\in T_\delta^n}\sqrt{p'_{\bk}}\ket{\bk \bk}^{AC}\ket{\varphi_{\bk}}^{BR}=\frac{1}{\sqrt{D_\delta^n}}\sum_{\bx'}\ket{\widetilde{\bx}'}^{A}\ket{\vartheta'_{\bx'}}^{CBR},
\end{align}
and this fact, coupled with the group covariance of both sets, gives us a means to transform $\Lambda^{CB}_{\alpha,\bx}$ into a measurement
$\Lambda'^{CB}_{\alpha,\bx'}$ suitable for distinguishing the $\vartheta'^{CB}_{\bx'}$.  

To see how this works, it is easiest to go back to the proof of the HSW theorem, which for convenience
is stated in the appendix. In the original i.i.d.\ case, the projectors $P_\bx$ and $P^{CB}$ onto the typical subspaces of $\vartheta_\bx^{CB}$ and
 $\bar{\vartheta}^{CB}=\frac{1}{2^n}\sum_\bx\vartheta_\bx^{CB}$, respectively, fulfill the five conditions needed in the proof of the theorem, equations~\ref{eq:c1} through~\ref{eq:l}. Since $\vartheta_\bx^{CB}=(Z^\bx)^C\Psi_0^{CB}(Z^\bx)^C$, the same holds for $P_\bx^{CB}$, and the five conditions
become 
\begin{align}
{\rm Tr}[\bar{\vartheta}^{CB}(\mathbbm{1}^{CB}-P^{CB})]&\leq \epsilon\label{eq:newc1}\\
{\rm Tr}[\Psi_0^{CB}(\mathbbm{1}^{CB}-P_0^{CB})]&\leq \epsilon\label{eq:newc2}\\
P_0^{CB}&\leq r\cdot\Psi_0^{CB}\label{eq:newr}\\
\sum_{\bx}\vartheta_\bx^{CB} &\leq d\cdot\bar{\vartheta}^{CB}\label{eq:newd}\\
||P^{CB}\bar{\vartheta}^{CB} P^{CB}||_\infty&\leq \lambda\label{eq:newl},
\end{align}
with $\epsilon=$, $r=2^{n[S(\psi^{CB})+\delta]}$, $d=2^n$ (and the condition is an equality since all $\bx$ are typical), $\lambda=2^{-n[S(\vartheta^{CB})-\delta]}$. 
Our aim is now to find a set of new projectors 
$P'^{CB}_\bx$ and $P'^{CB}$ fulfilling these conditions for the states $\vartheta'^{CB}_{\bx'}$ and $\bar{\vartheta}'^{CB}=\frac{1}{D_\delta^n}\sum_{\bx'}\vartheta'^{CB}_{\bx'}$. 

To start, use the fact that $ {\rm Tr}[({\Psi}'^{CB}_{{0}}-\Psi_{{0}}^{CB})P^{CB}_{\bf 0}]\leq ||{\Psi}'^{CB}_{{0}}-\Psi_{0}^{CB}||_1\leq \sqrt{\epsilon}$, 
since the trace distance is equal to the maximum of the lefthand side, maximized over all projectors~\cite{nielsen_quantum_2000}. Then we have 
\begin{align*}
{\rm Tr}\left[(\mathbbm{1}-P^{CB}_{\bf 0}){\Psi}'^{CB}_{{0}}\right]\leq{\rm Tr}\left[(\mathbbm{1}-P^{CB}_{\bf 0}){\Psi}^{CB}_{{0}}\right]+ ||{\Psi}'^{CB}_{{0}}-\Psi_{0}^{CB}||_1 \leq \epsilon+\sqrt{\epsilon},
\end{align*}
and so we can define $P^{CB}_{\bx'}=(Z'^\bx)^C P^{CB}_{0}(Z^\bx)^{C}$ to satisfy the first condition. 
The second condition follows analogously upon noting that 
$\bar{\vartheta}^{CB}=\sum_{{\bk}}p_{{\bk}} \ketbra{\bk}^{C}\otimes \varphi_{{\bk}}^B$ (and similarly for the pruned version) and 
therefore $||\bar{\vartheta}'-\bar{\vartheta}||\leq 2(1-\mathcal{N}_\delta^n)\leq 2\epsilon$.  
The third condition remains as is, since we're using the same $P_{\bf 0}$, and the fourth is an equality when $d=D_\delta^n$. For the fifth condition, observe that 
\begin{align*}
\tfrac{1}{\mathcal{N}_\delta^n}\bar{\vartheta}^{CB}-\bar{\vartheta}'^{CB}&=\tfrac{1}{\mathcal{N}_\delta^n}\sum_{\bk\notin T_\delta^n} p_\bk \ketbra{\bk}^C\otimes \varphi_\bk^B\geq 0.
\end{align*}
Therefore, $P^{CB}\bar{\vartheta}'^{CB}P^{CB}\leq \frac{1}{\mathcal{N}_\delta^n}P^{CB}\bar{\vartheta}^{CB}P^{CB}$, which leads immediately to  
$||P^{CB}\bar{\vartheta}'^{CB}P^{CB}||_\infty\leq \lambda/\mathcal{N}_\delta^n\leq \lambda(1+2\epsilon)$. 

We thus have all the ingredients needed to construct the
required measurement, with $\epsilon'=2\sqrt{\epsilon}$, $r'=r$, $d'=D_\delta^n$, and $\lambda'=\lambda(1+2\epsilon)$. 
 The number of syndromes Bob needs from Alice is given by 
$m_X'\geq n[S(\psi^{AB})+S(\psi^A)-S(\vartheta^{CB})+3\delta]+\log(1+2\epsilon)$, which works out to be 
$m_X'\approx n[S(\psi^R)-\sum_k p_kS(\varphi_k^R)]$. 
Since the pruned state is nearly identical to the original state, the remainder of the protocol 
goes through as before, outputting roughly $nS(A)-m_Z-m_X'$ entangled pairs. A simple calculation (along the lines of lemma 2 
in~\cite{renes_physical_2008}) gives $m_X'+m_Z=I(A{:}E)$ and $nS(A)-m_X'-m_Z=-S(A|B)$, and thus the protocol is optimal.

\section{\label{sec:summary} Conclusion}
We have shown how to construct an optimal state merging protocol by following the intuition from quantum error-correction 
that what really matters in two-party quantum information processing is 
information about amplitude and phase measurements. Combining entanglement distillation with teleportation, our results also imply a new proof of the direct part of the noisy channel coding theorem~\cite{devetak_distillation_2005}, one not following the usual route of decoupling Alice's system from the purification $R$ (e.g. all the fully fleshed-out proofs to date~\cite{devetak_private_2005,hayden_decoupling_2008,klesse_random_2008,horodecki_quantum_2008,hayden_random_2008}). It would be interesting to apply these techniques to more protocols, and see how far this intuition about quantum information extends.

\noindent {\bf Acknowledgments}
JMR received support from the European IST project SECOQC and JCB from Quantumworks and the Natural Sciences and Engineering Research Council of Canada (NSERC).
\appendix

\section{Static HSW Theorem}
Here we are interested in the ``static'' setting of the HSW theorem, which is concerned with the following.  
Given $n$ samples from an ensemble $\{p_k,\rho_k\}_{k=1}^d$ with average $\rho=\sum_k p_k\rho_k$, what is the smallest amount of 
side information $t=f(\bk)$ required in order to reliably construct a measurement $\Lambda_{t,\bk}$ which will identify $\bk$ from $\rho_\bk$ with only a small probability of error? In order to match the setting in the main text, we can think of the ensemble as arising from the state $\psi^{AB}=\sum_k p_k\ketbra{k}^A\otimes \rho_k^B$, a measurement of $\ket{k}$ (or $Z^A$) on $A$ generating state $\rho_k$. 
For random CSS codes $f$ is a random linear function, resulting from
measuring the stabilizer observables on the state $\ket{\bk}$. However, in what follows we will consider \emph{universal hashing}~\cite{carter_universal_1979}, since it is no more difficult to do so. In universal (or 2-universal) hashing, the function  $f:\{0,1\}^n\rightarrow \{0,1\}^m$  generating the side information is chosen at random from a universal family of hash functions in which the probability of collision $f(x)=f(y)$ but $x\neq y$ is the same as for random functions: Pr$_f[f(x)=f(y)|x\neq y]\leq 1/2^m$.

In~\cite{renes_physical_2008} we proved that for a fixed $\delta>0$, choosing $m=n[S(Z^A|B)+4\delta]$ is sufficient to guarantee 
the existence of a measurement having elements $\Lambda_{f(\bk),\ell}$ such that the probability of error P$_e$ is exponentially small:
\begin{align}
{\rm P}_e=\bigg\langle \sum_{\ell\neq \bk}\left[\Lambda_{f(\bk),\ell}\rho_\bk\right]\bigg\rangle_{f,\bk}\leq 6\times 2^{-n\delta^2/2}.
\end{align}
A crucial step in the proof is to show the existence of projectors $Q_\bk$ and $Q$ such that 
\begin{align}
{\rm Tr}[\left\langle\rho_\bk\right\rangle_\bk(\mathbbm{1}-Q)]&\leq \epsilon\label{eq:c1}\\
\left\langle{\rm Tr}[\rho_\bk(\mathbbm{1}-Q_\bk)]\right\rangle_\bk&\leq \epsilon\label{eq:c2}\\
Q_\bk&\leq r\cdot\rho_\bk\label{eq:r}\\
\sum_{k\in T_\delta^n}\rho_\bk &\leq d\cdot\left\langle\rho_\bk\right\rangle_\bk\label{eq:d}\\
||Q\left\langle\rho_\bk\right\rangle_\bk Q||_\infty&\leq \lambda\label{eq:l},
\end{align}
after which it can be shown that $m\geq \lfloor \frac{1}{\gamma}\log rd\lambda\rfloor$ for $0\leq \gamma\leq 1$ suffices to construct the 
measurement.\footnote{Breaking up the proof in this way is similar to the packing lemma of~\cite{hsieh_entanglement-assisted_2008}.} In the i.i.d.\ case of the HSW theorem, 
the $Q_\bk$ and $Q$ are projectors onto the typical subspaces of $\rho_\bk$ (for typical $\bk$) and $\rho^{\otimes n}$, respectively,
for which $\epsilon=2^{-cn\delta^2}$, $r=2^{n[\sum_k p_kS(\rho_k)+\delta]}$, $d=2^{n[H(p_k)+\delta]}$, and $\lambda=2^{-n[S(\rho)-\delta]}$. Thus, one chooses 
$m\geq n[H(p_k)-S(\rho)+\sum_k p_kS(\rho_k)+4\delta]=n[S(Z^A|B)+4\delta]$.


\end{document}